\documentclass[conference]{IEEEtran}
\IEEEoverridecommandlockouts
\usepackage{cite}
\usepackage{amsmath,amssymb,amsfonts}
\usepackage{algorithmic}
\usepackage{graphicx}
\usepackage{textcomp}
\usepackage{xcolor}

\usepackage{relsize}

\def\BibTeX{{\rm B\kern-.05em{\sc i\kern-.025em b}\kern-.08em
    T\kern-.1667em\lower.7ex\hbox{E}\kern-.125emX}}

\begin{document}



\title{Adapting The Secretary Hiring Problem for Optimal Hot-Cold Tier Placement under Top-$K$ Workloads
\thanks{
[Affliations and acknowledgments blinded for review]
The HASTE Project (http://haste.research.it.uu.se/) is funded by the Swedish Foundation for Strategic Research (SSF) under award no. BD15-0008, and the eSSENCE strategic collaboration for eScience.
}
}

\author{
\IEEEauthorblockN{
Ben Blamey\IEEEauthorrefmark{1}, 
Fredrik Wrede\IEEEauthorrefmark{1}, 
Johan Karlsson\IEEEauthorrefmark{2},
Andreas Hellander\IEEEauthorrefmark{1},
and Salman Toor\IEEEauthorrefmark{1}
}
\IEEEauthorblockA{
\IEEEauthorrefmark{1} Department of Information Technology, Division of Scientific Computing, Uppsala University, Sweden\\ 
Email: \{Ben.Blamey, Fredrik.Wrede, Andreas.Hellander, Salman.Toor\}@it.uu.se 
}

\IEEEauthorrefmark{2} Discovery Sciences, Innovative Medicines, AstraZeneca, Gothenburg, Sweden\\ 
Email: johan.karlsson1@astrazeneca.com
}



\maketitle

\begin{abstract}
Top-K queries are an established heuristic in information retrieval. This paper presents an approach for optimal tiered storage allocation under stream processing workloads using this heuristic: those requiring the analysis of only the top-$K$ ranked most relevant, or most interesting, documents from a fixed-length stream, stream window, or batch job. In this workflow, documents are analyzed for relevance with a user-specified \emph{interestingness function}, on which they are ranked, the top-$K$ being selected (and hence stored) for further processing. This workflow allows \emph{human in the loop} systems, including supervised machine learning, to prioritize documents. This scenario bears similarity to the classic Secretary Hiring Problem (SHP), and the expected rate of document writes, and document lifetime, can be modelled as a function of document index.
We present parameter-based algorithms for storage tier placement, minimizing document storage and transport costs.
We show that optimal parameter values are a function of these costs. It is possible to model application IO characteristics analytically for this class of workloads. When combined with tiered storage, the tractability of the probabilistic model of IO makes it possible to optimize (and budget for) storage tier allocation \emph{a priori}, without needing to monitor the application. This contrasts with (often complex) existing work on tiered storage optimization, which is either tightly coupled to specific use cases, or requires active monitoring of application IO load (a reactive approach) -- ill-suited to long-running or one-off operations common in the scientific computing domain. We evaluate our model with a trace-driven simulation of a bio-chemical model exploration, and give case studies for two cloud storage case studies.
\end{abstract}


\section{INTRODUCTION}
Selecting the top-$K$ documents ranked according a scoring function is known as a \emph{top-$K$ query}. In this paper, we present a strategy for optimal use of tiered storage during the computation of these queries over large sequences, or streams, of documents. We show how the \emph{Secretary Hiring Problem} can be adapted to reduce storage and communication costs where only the most relevant (or interesting) ranked documents are analyzed at the end of a batch, fixed-length stream, or stream window. 

This approach ideally suited to workflows where it is only possible to process a subset of input data, for example, human in the loop (HITL) systems, including supervised or semi-supervised machine-learning~\cite{wredeSmartComputationalExploration2018}, or active learning; as well as computational bottlenecks.
In the case of stream processing applications, if the data input rate consistently exceeds the processing rate, some data must be removed from the processing queue, if latency is not to diverge. Unprocessed data can be archived or discarded. The top-$K$ heuristic is a sensible approach to such load shedding when resources are insufficient for processing the full stream. There is precedent for load shedding based based on relevance or importance to the context~\cite{tatbulLoadSheddingData2003}. 

Regardless of processing bottlenecks, some workflows (especially in the scientific computing domain) can  generate more data than is economically feasible to store. Retaining only the top-$K$ most interesting or relevant documents is a reasonable approach.
In this paper,
we assume a user-provided \emph{interestingness function}\footnote{c.f. the notion of \emph{scoring function} in much of the information retrieval literature.} for document ranking, and consider optimal use of tiered storage under a generalized stream processing workflow. 

A stream of length $N$ is equivalent to repeated stream analysis workload with a non-overlapping window of length $N$, treating each window independently. 





\section{MOTIVATION}
As discussed, top-$K$ workflows are a useful approach for making best use of constrained processing (human or computstional) and storage resources -- in the face of large datasets. In this section, we justify an interest in tiered storage in this context, and by implication, optimal tiered storage management. 

In conventional enterprise stream processing applications, message sizes are perhaps a few KB (JSON, XML documents). Message processing frequencies may approach millions of documents a second in larger systems. 
For these systems, reducing latency, and increasing maximum frequency throughput are key research goals.
In scientific computing, and many video and imaging related applications, message sizes can be much larger. 
Spatial cell simulations are often in the range 1MB-100MB (parameter sweeps can generate millions of such documents), and microscopy images are typically a few MB each.
Cost, rather than absolute performance, can be more a driver. These massive workloads~\cite{deelmanDataManagementChallenges2008}, are different from enterprise streaming workloads, as is the economic context. In enterprise contexts, perceived data value motivates a ``keep everything'' strategy; whereas in a scientific context, this can be economically infeasible.
Driven by these considerations, recent work explores more creative and unconventional use of cloud computing resources. For example, \cite{shankarNumpywrenServerlessLinear2018} is a system for linear algebra on large matrices, which eschews traditional VMs and storage in RAM for object storage and FaaS. 

In addition to large messages, unusually long stream processing windows create a requirement for longer-term storage of intermediate data. In industrial research settings, microscopes can generate many hours of high-resolution imagery.

Tiered, hybrid or hierarchical storage systems typically combine a number of tiers of more expensive, high-performance media with cheaper low-performance storage. This presents an opportunity for cost savings on workloads requiring storage of large numbers of large documents for extended timeframes. 

\section{RELATED WORK}
This paper draws on three areas of literature: (a) work on top-K queries, (b) work on optimizing the use of tiered and hybrid storage; and (c) various classic discrete optimization problems which serve as their basis. 

Top-$K$ queries are well studied within the information retrieval and database communities, making them a well-established heuristic for working with large datasets. See~\cite{ilyasSurveyTopkQuery2008} for a survey. Two research challenges are improving query optimization in the context of distributed systems~\cite{caoEfficientTopkQuery2004}, and streams \cite{babcockDistributedTopkMonitoring2003a}. Whilst this paper assumes the implementation of top-$K$ queries, our concern is instead with storage tier allocation of large documents returned by such a query, and the associated storage/transport cost-optimization problems. Note that the discovery of the top-$K$ \emph{most frequent} documents~\cite{charikarFindingFrequentItems2002} is a distinct problem, outside the scope of this paper.

There is a significant body of existing work relating to hybrid, hierarchical, hot/cold and tiered storage. Broadly, the goal of this work has always been to utilize different storage devices and media, with different performance characteristics and costs, to maximize read/write performance and/or minimize costs. As new storage media (flash and SSDs) have become available, and the way we think about and pay for storage has changed (cloud computing), the basis of these optimization problems have evolved, as have the frameworks and systems to work with tiered storage.

Much of this existing work could be characterized \emph{as
reactive}, based on low-level file metadata available in that filesystem (file creation time/age, number; frequency and frequency of file access) -- without high-level application metadata.

Early work on hybrid storage concerned management of tiers of high performance and archive disks and tape drives, respectively. In \cite{wilkesHPAutoRAIDHierarchical1996} tiers were built from different RAID configurations -- and tier placement was reactive: ``[migrating] data blocks between these two levels as access patterns change". In a similar vein, continuous Markov Chain models were used in \cite{kraissIntegratedDocumentCaching1998}, transition probabilities being estimated ``through access monitoring''.

Later, the large datasets in video and multimedia-streaming services motivated access models (and tier placement strategies) specific to that context. \cite{sapinoLoganalysisBasedCharacterization2006} adopted a reactive approach -- building finite state automata from access logs, whereas \cite{chanModelingDimensioningHierarchical2003} modelled document read frequency as a Zipf distribution. The notion of document age as a predictor of document heat (and hence tier), is a popular heuristic, nicely exemplified in \cite{muralidharF4FacebookWarm2014}. 

These techniques are not limited to hybrid storage. \cite{xiaFARMERNovelApproach2008} improves on previous work which uses ``file access sequence and semantic attribute'', by mining correlations between documents. \cite{olyMarkovModelPrediction2002} used application traces to form a Markov model to predict block access. Similarly in  \cite{dorierOmniscIOGrammarBased2014}: ``modeling the I/O behavior of any HPC application
and using this model to accurately predict the spatial and
temporal characteristics of future I/O operations''.

The growing popularity of SSDs motivated extensive work on hybrid HDD/SSD tier management systems, with awareness of IO characteristics of both media (e.g. random vs sequential writes). See \cite{niuHybridStorageSystems2018} for a review. In this work, there is a strong emphasis on caching heuristics  -- using document age and access frequency as a predictor of future access; i.e. a reactive approach. More elaborate techniques are possible using information available from the filesystem: \cite{liCMinerMiningBlock2004} mined correlations between blocks, allowing read-ahead optimization.
More recently, \cite{krishEfficientHierarchicalStorage2016} present a system which monitors the IO workload of popular big-data processing frameworks, allocating data between HDD and SSDs accordingly. 
Broadly, performance improvement (constrained by cost) is the optimization objective of these systems, investment in storage tiers being made upfront. 

The flexibility of cloud computing removed the need for upfront investment,
and allowed optimization based primarily on reducing cost (subject to performance constraints), especially in the case of object storage and similar long-term storage platforms. The simple pricing models, multitude of available tiers and providers led to extensive work on optimized cloud storage tier placement. 

\cite{abu-libdehRACSCaseCloud2010} suggest RAID-style striping and replication across different CSPs for resilience and to reduce costs. In the Frugal Cloud File System \cite{puttaswamyFrugalStorageCloud2012}, workloads are analyzed to adapt the size of an EBS working set, flushing to S3 (c.f. Case Study 2, section~\ref{sec:casestudy2}). The system is reactive, assuming that ``we do not have any knowledge of when the next access will occur''. \cite{dealBudgetTransferLowCost2018} move data between providers to minimize long term costs. 

Many of these systems rely on the input of quantified access and performance requirements; which in practice can be difficult to formulate for real-world systems. Frameworks such as  \cite{raghavanTieraFlexibleMultitiered2014, kakoulliOctopusFSDistributedFile2017} to manage data over multiple media, tiers and providers, according to a given input policy.


There is precedent for applying discrete optimization problems to cloud computing. \cite{khanaferConstrainedSkiRentalProblem2013a} apply the ski-rental problem -- first introduced for TCP caching \cite{karlinCompetitiveRandomizedAlgorithms1990} -- to make an optimal tradeoff between the cost of caching a web document, and re-computing it. With a storage focus, \cite{mansouriCostOptimizationAlgorithms2018}, used dynamic programming to express represent strategies for optimal tier placement, applying the \emph{ski rental} problem. This work allows proactive tier placement, but information about the file operations still needs to be provided in a form suitable for their optimization approach.

The novelty of our work is our specific focus on document storage (and transport) for top-$K$ workloads: adapting the secretary hiring problem yields strategies with closed-form expressions (in terms of costs) for optimal parameter values. By modelling the workload in this way, we can predict the expected number of document reads and writes depending on stream position; rendering the optimal placement strategy tractable -- and allowing proactive tier placement (without real time monitoring). Whilst focusing on a particular class of workloads, our approach contrasts with the approach in much of the literature. To the authors' knowledge, there is no previous work on document storage for top K workloads.

\section{THE OPTIMIZATION PROBLEM}
\begin{figure*}
\includegraphics[clip, trim=0cm 4.5cm 0.3cm 0.3cm, width=1.00\textwidth]{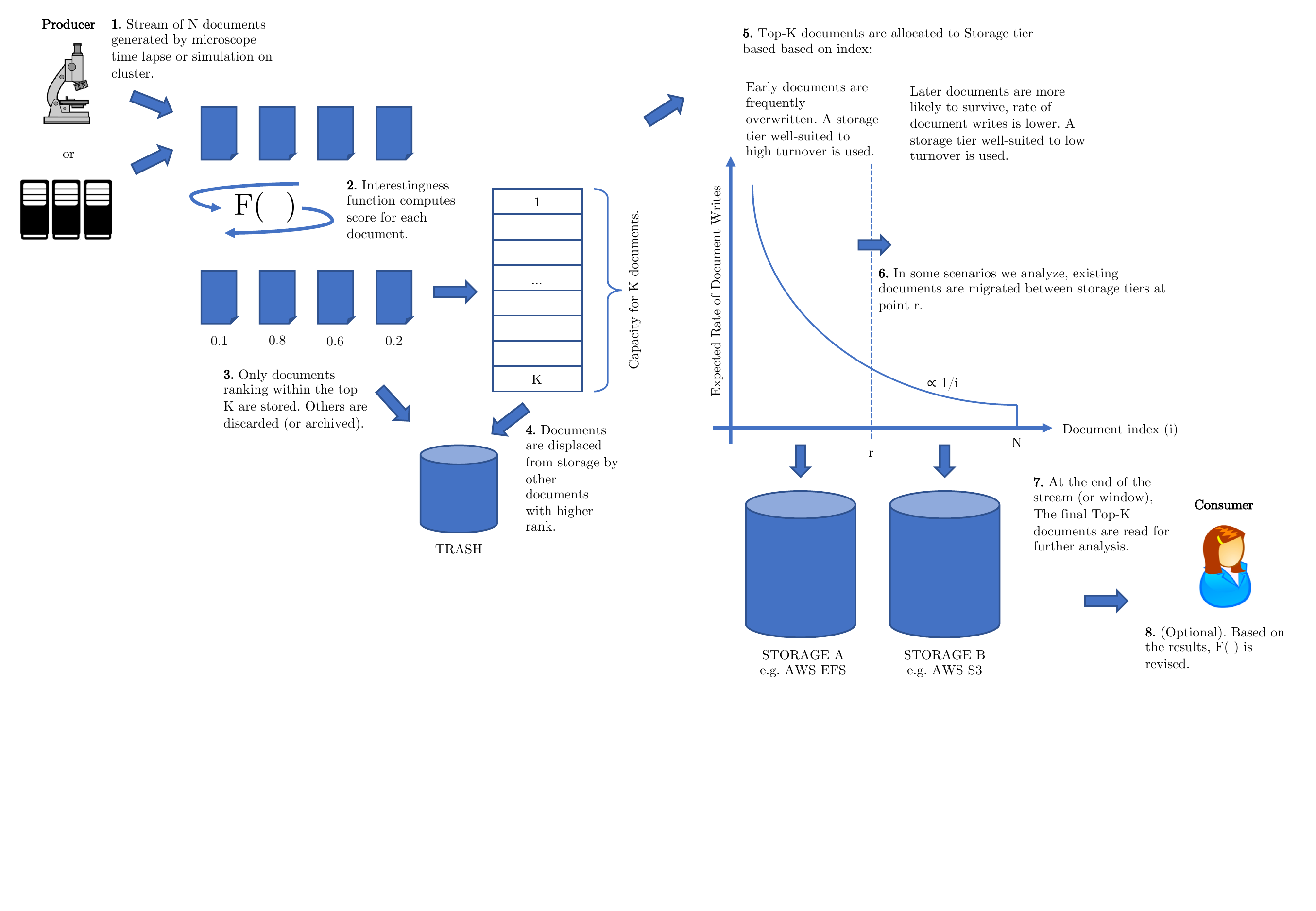}
\caption{\label{fig:arch}Our Top-$K$ Stream Processing Architecture with 2 abstract Storage Tiers. Storage Tiers A and B could represent 
storage which is local to the producer and consumer respectively in a multi-cloud context (Case Study 1), where communication costs are accounted in the write and read costs respectively for the two tiers. Or, they could be simply be different storage tiers in the same cloud (e.g. AWS EFS and S3, as in Case Study 2.)}
\end{figure*}

The architecture under study (Fig.~\ref{fig:arch}) includes a producer and a consumer, pre-processing with a user-provided \emph{interestingness function}, and two abstract storage tiers. Read and write costs are modelled independently for the two storage tiers. Transport costs can be accounted for in the storage tier read and write costs (in cloud/edge and multi-cloud cases). 
The consumer read the documents, and performs further analysis on the final top-$K$ documents in the source stream, beyond the scope of this paper.

Central to our approach is a user-provided interestingness function   
able to compute the relevance of the input documents. This could for example be a pre-trained classifier or regressor that, based on features derived from the documents, predict the likelihood of a document being prioritized for downstream analysis. 
Under the top-$K$ workflow, interestingness translates to the read probability according to a simple probabilistic model.
Our algorithm assigns each document to a storage tier given the interestingness scores, online. 
This work is equally applicable to streams of data originating in the cloud (such as simulation output), as well as data from other sensors (such as microscopes) connected at the cloud edge.



A stream of documents is produced (e.g. by a microscope, or simulation running in a cluster), and analyzed for relevance by an interestingness function. 
Each document is ranked in turn against those already produced on the output of this function, an \emph{interestingness score}.

After the production of a fixed-length stream of $N$ documents, the top-$K$ most interesting will be processed. The documents must be stored in the meantime, and possibly transferred to another location for subsequent analysis (requiring that we account for transfer costs in the optimization). 
Clearly, as the stream progresses, the set of $K$ most interesting documents will change, as more interesting documents are observed and stored, overwriting less-interesting documents.

More formally:
consider a sequence $d_i$ of $N$ documents, from a stream $D$. An interestingness function $H(d_i)$, gives $h_i$, the document interestingness, inducing a rank on $D$. We suppose $D$ and the values of $h_i(d)$ over $D$ are unknown \emph{a priori}. To model the cost of the reads, we assume that the top-$K$ documents will be read by the consumer after all documents have been produced.

\section{ALGORITHM A: The Classic Secretary Hiring Problem}
First, we recap the classic secretary hiring problem.
There are $N$ ranked job candidates, which are interviewed in turn. After each interview, the rank of the candidate against previous candidates is known. You can decide to (irrevocably) hire that candidate, or continue interviewing. You only know the rank of the candidates you have interviewed so far. How can you maximize the probability that you hire the best overall candidate? It can be shown that the optimal strategy is as follows \cite{dynkin1963optimum}: for given parameter $r$; observe the first $r - 1$ candidates, and hire the next candidate who ranks higher than the best candidate $M$ among the first $r-1$. The optimal value of $r$ is actually a function of $e$, the base of natural logarithms:

\begin{align}
N & = \textrm{Number of documents} \\
r_{\textrm{optimal}} & = \frac{N}{e} \\
P(\textrm{hiring overall best}) & = \frac{1}{e} = 0.367  \\
\mathbb{E}(\textrm{\# writes}) & = 1
\end{align}

\section{ALGORITHM B: \emph{Simple Overwrite}, $K=1$, 1 tier}

\begin{figure}
\begin{small}
\begin{verbatim}
Require: 
N, K

H = [] # Sorted list
for d_i in D:
    h_i = interestingess(d_i)
    H.insert(h_i)
    h_rank = H.indexof(h_i)
    if h_rank < K:
        store(d_i, Storage_A)
        # Retain only the top K docs: 
        prune(Storage_A)
\end{verbatim}
\end{small}
\caption{Simple Overwrite, 1 tier, $K>1$.}
\label{alg:simple-one-tier}
\end{figure}

We can adapt the secretary hiring problem to tiered storage by considering documents as candidates, ranked according to an interestingness function. Furthermore, in our scenario, previous `best' documents can be overwritten by the producer to ensure that the best overall are always available for further analysis. This differs from the classic SHP problem. First consider a simple example with 1 tier of capacity 1, and simply compute the expected number of writes for a stream of $N$ documents under the algorithm listed in Fig.~\ref{alg:simple-one-tier}. This is a simple starting example to understand the document overwrite probabilities, there is no $r$ parameter.

\begin{align}
P(\text{$i$th document is best so far}) & = \frac{1}{i + 1} \\ 
\mathbb{E}(\text{\# writes}) & =  \mathlarger{\mathlarger{\sum}}_{i=0}^{N-1} \frac{1}{i+1}
\end{align}
This partial sum of the Harmonic series can be approximated:
\begin{align}
\mathbb{E}(\text{\# writes}) & = \ln{N} + 0.57722 \\
P(\text{saving best}) & = 1
\end{align}

Compare with the solutions to the classic secretary hiring problems above. It is trivial to extend this approach to $K > 1$. Note this is unrelated to the solution for $K>1$ for the classic (`hiring-only') problem (without replacement), using a recursive approach \cite{kleinbergMultipleChoiceSecretaryAlgorithm2015}.

\section{ALGORITHM C: \emph{``First r to A, the rest to B''}, k $>$ 1, 2 tiers} \label{sec:algC}
If we extend algorithm B to more than 1 tier, we have a new optimization problem -- a decision about \emph{where} to save a `best-so-far' document to reduce overall costs.

Intuitively, at the start of the stream, `best-so-far' documents are likely to be overwritten (and unlikely to survive until the read): such documents could be written to write-optimized, or producer-local storage. Conversely, towards the end of the stream, intermediate-best documents are less likely to be overwritten, and more likely to survive and be read. Under these scenarios we can compute optimal `changeover' strategies for high-performance and low-performance storage (drawing inspiration from the optimal classic SHP strategy), as well as hedging transfer costs against storage transaction costs. We can adapt the algorithm from the classic secretary hiring problem to the 2-tier case, see fig.~\ref{alg:simple-two-tier}. From there, we can derive expressions for the expected number of reads, and the associated costs (assuming $r > k$):
\begin{align}
P\Bigg(\text{\parbox{4cm}{$i$th doc. in top $K$ when observed}} \, | \, i < K\Bigg) = & \, 1 \label{eq-algC-rateI}
\end{align}
\begin{align}
P\Bigg(\text{\parbox{4cm}{$i$th doc. in top $K$ when observed}} \, | \, i \geq K\Bigg) = & \, \frac{K}{i + 1} \label{eq-algC-rateII}
\end{align}
\begin{align}
\mathbb{E}\Bigg(\text{\parbox{4cm}{cumulative number of writes }}| \, i < K\Bigg) = & \, i \label{eq-algC-cumw-I}
\end{align}
\begin{align}
\mathbb{E}\Bigg(\text{\parbox{2cm}{cumulative number of writes }}| \, i \geq K\Bigg) = & \, K + K \cdot \ln{(i + 1)} \label{eq-algC-cumw-II}
\end{align}
\begin{align}
\mathbb{E}(\Sigma\textrm{ cost writes}) = & \, \mathlarger{\mathlarger{\sum}}_{i=0}^{N-1} P\Bigg( \text{\parbox{2.5cm}{$i$th document is in top $K$ when observed}}\Bigg) \notag \\ & \,\, \, \, \cdot (\text{cost of writing it})
\end{align}
\begin{align}
\mathbb{E}(\Sigma\textrm{ cost writes}) = & \, K \cdot cost_{\textrm{write, A}} \notag\\
+ & \,\,\, K \cdot ( \ln{r} - \ln K ) \cdot cost_{\textrm{write, A}}\notag\\
+ & \,\,\, K \cdot ( \ln{N} - \ln {r} ) \cdot cost_{\textrm{write, B}} \label{eq-writes}
\end{align}

Expressions (\ref{eq-algC-rateI}) and (\ref{eq-algC-rateII}) are equivalent to the expected rate (per document) of document writes. 
That is, 
assuming that the documents are randomly ordered w.r.t the rank of their interestingness score, we see that the rate of document overwrite (i.e. write and transfer cost) is $K/(i + 1)$ ($i = K,..,N-1$) 
\footnote{The first $K$ documents will be always be written.} whereas document storage costs are a linear function of time. 

\begin{figure}
\begin{center}
\includegraphics[width=0.4\textwidth]{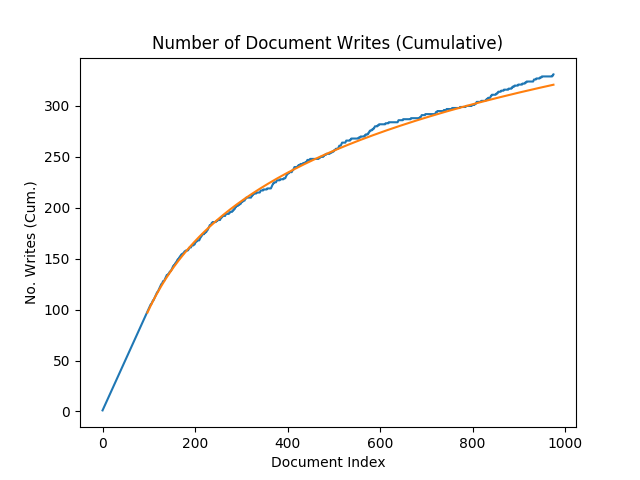}
\end{center}
\caption{Number of Document Writes (Cumulative), for interestingness trace of gene regulatory network simulations. The analytic solution, i.e. equations (\ref{eq-algC-cumw-I}) and (\ref{eq-algC-cumw-II}), are shown for comparison. We model of the total number of writes to make an \emph{a priori} decision on storage tier selection. Notice that the first $K=100$ documents are all written. }

\label{fig:cum-writes}
\end{figure}

Equations (\ref{eq-algC-cumw-I}) and (\ref{eq-algC-cumw-II}) are shown on fig.~\ref{fig:cum-writes}, plotted against a real trace from a case study discussed in Section VIII.



\begin{figure}[h]
\begin{small}
\begin{verbatim}
Require: 
N, K, r, DO_MIGRATE

H = [] # Sorted list
for d_i in D:
    h_i = interestingess(d_i)
    H.insert(h_i)
    h_rank = H.indexof(h_i)
    
    if i == r AND DO_MIGRATE:
        # Migrate all docs.
        for d_j in Storage_A:
            migrate(d_j, Storage_A, Storage_B)
    
    if h_rank < K:
        if i < r:
            store(d_i, Storage_A)
        else:
            store(d_i, Storage_B)
            
        # Retain only the top K docs:
        prune(Storage_A)
        prune(Storage_B)
\end{verbatim}
\end{small}
\caption{\label{alg:simple-two-tier}\emph{``First r to A, the rest to B''}, $K>1$, 2 tiers}
\label{alg:c}
\end{figure}

In cases where transaction and communication costs dominate, documents remain in the tier in which they are written (i.e. no migration\footnote{i.e. \texttt{DO\_MIGRATE = False} in Fig.~\ref{alg:c}.}). 
Without migration, the number of documents stored in each platform would change throughout the stream. 
If rental costs are relatively small, it is simpler to use a bound (using the most expensive tier rental). Using this bound means rental costs are constant (in $r$).
By introducing this bounded approximation, we need to check that the 2-tier changeover strategy has lower overall costs than simply using one of the tiers exclusively.
Supposing the read operation occurs after the stream, we can assume the best overall documents are i.u.d. over the stream:
\vspace{-3mm}

\begin{align}
\mathbb{E}(\Sigma \, \textrm{cost reads})  = & \, K \cdot \Big( \frac{r}{N}
\cdot cost_{\text{read, A}} \notag \\ 
& + (1 - \frac{r}{N}) \cdot cost_{\textrm{read, B}} \Big) \\
\mathbb{E}(\Sigma \, \textrm{cost total})  = & \, \mathbb{E}(\Sigma \textrm{cost writes})  + \mathbb{E}(\Sigma \, \textrm{cost reads})
\end{align}

\noindent
We can differentiate (w.r.t. $r$), and set to zero to find $r_{optimal}$:
\vspace{-4mm}

\begin{align}
0 = & \, \frac{K}{r_{\textrm{optimal}}}(cost_{\textrm{write, A}} - cost_{\textrm{write, B}}) \notag \\ & \, \, + \frac{K}{N}(cost_{\textrm{read, A}} - cost_{\textrm{read, B}}) \\
\frac{r_{\textrm{optimal}}}{N} = & \, \frac{cost_{\textrm{write, A}} - cost_{\textrm{write, B}}}{cost_{\textrm{read, B}} - cost_{\textrm{read, A}}}
\end{align}

Where storage rental costs are more considerable, it can be cheaper to migrate all documents. If documents are migrated when $i = r$ (incurring a read/write cost independent of $r$), then the storage rental costs are a linear function of $r$. 
The costs of the migration, and final read are constant in $r$:
\vspace{-4mm}

\begin{align}
\mathbb{E}(\Sigma \, \textrm{cost rental}) = & \, K \cdot \Big(\frac{r}{N} \cdot cost_{\text{rental, A}} \notag \\ & \, + (1 - \frac{r}{N}) \cdot cost_{\textrm{rental, B}} \Big) \\
(\textrm{cost migration}) = & K \cdot (cost_{\textrm{read, A}} + cost_{\textrm{write, B}}) \\
\mathbb{E}(\Sigma \, \textrm{cost total})  = & \, \mathbb{E}(\Sigma \, \textrm{cost writes})  + \mathbb{E}(\Sigma \, \textrm{cost rental}) \notag \\
& \, \, \, + (\textrm{cost migration})
\end{align}
\noindent

As before, we can differentiate (w.r.t. $r$), and set to zero to find $r_{optimal}$:
\vspace{-2mm}

\begin{align}
\frac{r_{\textrm{optimal}}}{N} = \frac{cost_{\textrm{write, A}} - cost_{\textrm{write, B}}}{cost_{\textrm{rental, B}} - cost_{\textrm{rental, A}}}
\end{align}

Regardless of whether migration is performed, the approach is only valid if: $ K < r_{\textrm{optimal}} < N $. Fig.~\ref{fig:case2} shows an example of the expected total cost for varying $r$. Below, we apply this to real-world storage problems involving online analysis of the best (or most interesting) $K$ documents in a stream.




\subsection{Case Study 1: 2 Tiers in Different Clouds}

Our first example is motivated by IoT/fog and multi-cloud computing scenarios. We here assume that the producer and consumer are geographically separated and situated with different cloud providers, or that the producer is at the fog/edge. Consumer-local storage has lower write transaction costs than producer local storage, yet producer and consumer are separated by a costly (or otherwise constrained) communication channel. At the start of the stream, documents are less likely to survive; intermediate documents may be better stored at the producer -- and \emph{pulled} to the consumer if it survives. Towards the end of the stream (window) the document is more likely to survive and be read. It may be cheaper to \emph{push} the document to the consumer-local storage. For example, suppose that the data is generated at an AWS cloud, where the S3 object store is available, and the consumer is situated in an Azure Cloud (where Azure Blob Storage storage is available). This scenario is illustrated in Table~\ref{table-case1}. All prices in USD\footnote{https://aws.amazon.com/s3/pricing/  - EU, Ireland (2018)}
\footnote{https://azure.microsoft.com/en-us/pricing/details/storage/blobs/ - GPv1 N. Eur.}
\footnote{https://azure.microsoft.com/en-us/pricing/details/bandwidth/ - N. Eur.}.

\begin{table}[ht]
\begin{tabular}{ l | p{2cm} }
\hline	
  N (Number of Documents) & 1e8 \\
  K & N/100 \\
  Document Size & 0.1 MB \\
  Stream Duration / Window & 1 day \\

  (A) Azure - PUT Cost & 0.00036 / 10,000 \newline= 3.6e-8 / doc. \\
  (A) Azure - GET Cost & 0.00036 / 10,000 \newline= 3.6e-8 / doc. \\
  (A) Azure - Rental Cost & 0.024 / GB month \\
    
  (B) S3 - PUT Cost & 0.005 / 1,000 \newline= 5e-6 / doc. \\
  (B) S3 - GET Cost & 0.0004 / 1,000 \newline= 4e-7 / doc. \\
  (B) S3 - Rental Cost & 0.023 / GB month \\
  
  Azure Transfer Costs (Out) & 0.087 / GB \\
  S3 Transfer Costs (In) & 0 / GB \\

$r_{opt}/N$ & 0.41233169 \\
\hline
Total Cost ($r_{opt}$, without migration, upper bound) & 35.19 \\
\hline
Other strategies... \\
Total Cost ($r_{opt}$, with migration) & 49.29 \\
Cost all Storage A: & 37.20 \\
Cost all Storage B: & 99.12 \\
\hline	
\end{tabular}
\caption{\label{table-case1}Price Breakdown, Case Study 1}
\end{table}

\begin{figure}[h]
\centering
\includegraphics[width=0.5\textwidth]{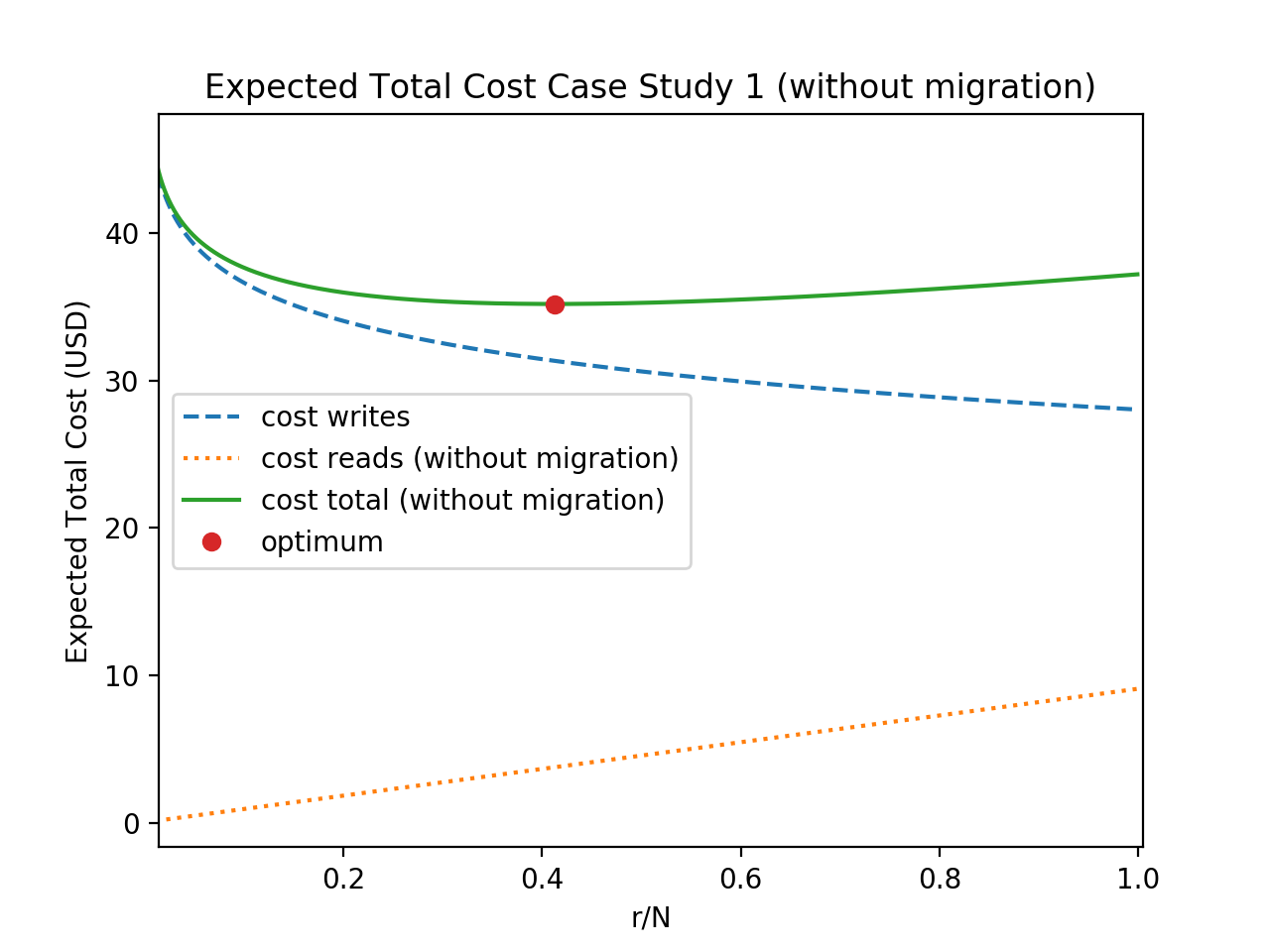}
\caption{\label{fig:case1}Expected overall cost for varying $r$, Case Study 1. Although the savings are perhaps marginal, they would accumulate over many window iterations.}
\end{figure}

Fig.~\ref{fig:case1} shows the value of $r$ which minimizes the expected total cost. The cost is minimized when the first 41\% of documents are saved in the producer-local storage, and the remaining documents saved to the consumer local storage.

\subsection{Case Study 2: 2 Tiers in the Same Cloud} \label{sec:casestudy2}

In this example, we are concerned with optimizing storage allocation within tiers in the same cloud provider or datacenter. This situation occurs frequently for scientific workflows where the application needs to balance use of more expensive but high-performing storage tiers with cheaper commodity storage. Generally, storage costs are modelled as a sum of 
read and write read and write costs
and \emph{rental costs} typically a product of document size and document lifetime. In cloud computing contexts, these costs are modelled explicitly. 
Under a top-$K$ processing workflow, the optimal choice of storage tier is a function of document index (and costs). At the start of the stream, when the rate of document replacement is high, storage tiers optimized for high-throughput are likely to be cheaper. Towards the end of the stream, document replacement (and hence read and write transactions) is lower, so storage tiers optimized for longer term storage may be more cost effective.

For example, consider two storage tiers in AWS: EFS, which is expensive to rent, but has low read and write costs, and the S3 object store. In this example, we include the rental costs. This scenario is illustrated in Table~\ref{table-case2}.


\begin{table}[ht]
\def\arraystretch{1.1}
\begin{tabular}{ l | p{2cm} }
\hline	
  N (Number of Documents) & 1e8 \\
  K & 5e6 (5\% of N) \\
  Document Size & 1 MB \\
  Stream Duration / Window & 7 days \\
  (A) EFS - Read & 0 \\
  (A) EFS - Write & 0 \\
  (A) EFS - Rental & 0.30 / GB Month \\ 

  (B) S3 - Read & 0.000005 \\
  (B) S3 - Write & 0.000005 \\
  (B) S3 - Rental & 0.023 / GB Month \\ 
  $r_{\textrm{optimal}}/N$ & 0.078 \\
\hline	
Total Cost ($r_{opt}$, with migration) $r_{opt}$ & 142.82 \\
\hline	
Other strategies... \\
Cost all Storage A & 350.00 \\
Cost all Storage B & 503.78 \\
Total Cost ($r_{opt}$, without migration, upper bound) & 415.67 \\
\end{tabular}
\caption{\label{table-case2}Price Breakdown, Case Study 2}
\end{table}
\noindent

\begin{figure}[h]
\centering
\includegraphics[width=0.5\textwidth]{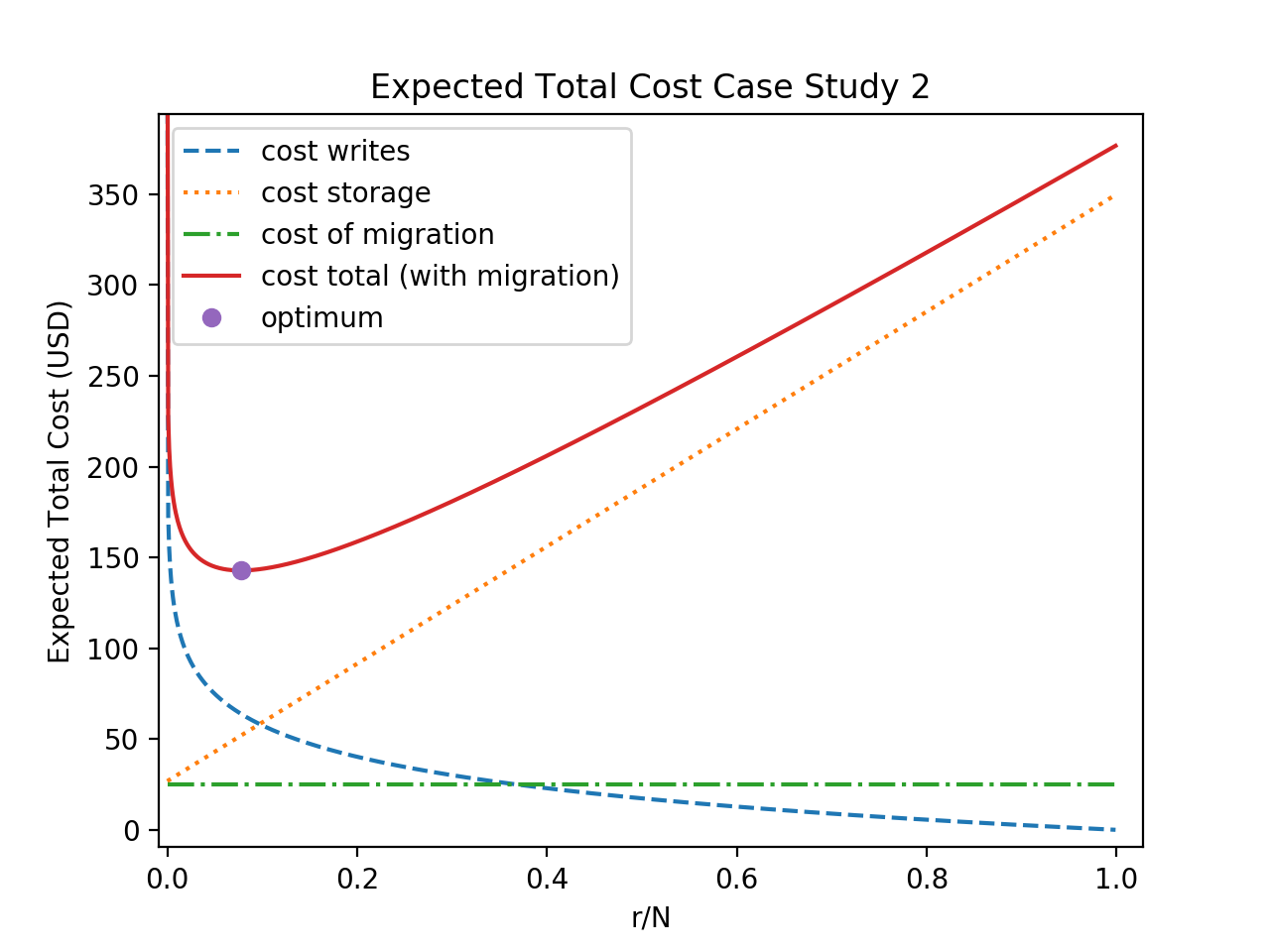}
\caption{\label{fig:case2}Expected overall cost for varying $r$, Case Study 2.}
\end{figure}

Fig.~\ref{fig:case2} shows the value of $r$ which minimizes the expected total cost. The expected cost is minimized the first 7.8\% of documents are saved in the producer-local storage (and pulled to the consumer during the migration), and the remaining documents pushed directly in the consumer local storage.

\section{EXAMPLE: A SEMI-SUPERVISED MODEL EXPLORATION WORKFLOW} \label{sec:sysbio}

The scenario in Case Study 2 in the previous section occurs frequently in cloud-based scientific computing workflows using elastic master-slave clusters. Typically, the cluster is equipped with a shared filesystem based on either instance storage or elastic block storage, in which both input data and simulation output can be written and accessed from compute nodes with low latency. 

An example of a computational experimentation application from computational systems biology that is organized this way is the MOLNs virtual cluster provisioning system for scalable cloud-native simulation of gene regulatory networks \cite{drawertMOLNsCloudPlatform2015}. MOLNs is part of the Stochastic Simulation Service, StochSS \cite{drawertStochasticSimulationService2016}. It 
exposes an API for depositing and accessing simulation output files either at a shared filesystem served from instance or block storage attached to the cluster master node, or in object storage. 

A common task is to conduct very large global parameter sweeps, where a stochastic biochemical model is simulated for large numbers of parameter combinations and the output of the sweep is subsequently analyzed in depth, to map out the range of possible variations in model behavior.

These documents are in the range of 0.1-100Mb for typical models. As the parameter space under exploration has large dimensionality, a significant number of messages are generated, resulting in very large datasets. 

\begin{figure}[t]
\includegraphics[width=0.5\textwidth]{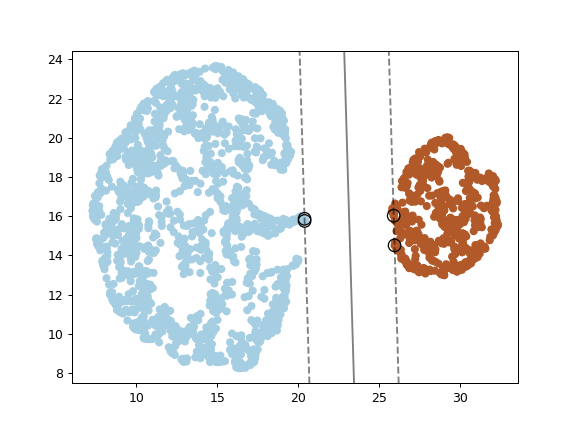}
\caption{Support vector machine used for interestingness evaluation of gene regulatory network simulations. Results falling into the left part of the figure (after feature extraction and dimension-reduction) are considered as non-interesting by the modeler while the ones to the right (red cluster) are considered interesting (in this case showing oscillatory behavior).}
\label{exp:svm}
\end{figure}

We used the tools from \cite{wredeSmartComputationalExploration2018}, combining smart sweeps and the herein proposed methodology, to construct a support vector machine (SVM) classifier, trained to classify simulation results from a model of a gene regulatory systems as biologically interesting or not (see fig.~\ref{exp:svm}). 

In such smart parameter sweeps the exploration proceeds in stages. First, with relatively few simulations and interactive human-in-the-loop active learning, the model is explored and the biologist informs the system about what simulation results should be considered interesting or not, and a classifier is trained. 
Then, very large datasets are generated by the simulators in a full sweep and stored for downstream analysis. In this analysis, it is most productive to focus on the outputs for which the classifier were most unsure (of either class), and again let the human modeler inform the system about those examples (active learning). Hence, an interestingness function can rank the documents according to the label entropy. 




As a measure of interestingness of the produced stream of simulation results, we used normalized label entropies. 
This approach to constructing an interestingness function is representative of many human-in-the-loop semi-supervised workflows.     
The underlying model in this example has 15 dimensions, with a uniform sampling of the parameter space, and a modest 10 independent samples of each parameter point, resulting in a computational experiment lasting for \textasciitilde$30$ days (with 60 cores), producing $143\times 10^6$ documents totalling $14.8$ TB of data. 



\section{CONCLUSIONS}

Adopting a top-$K$ workflow on the basis of a user-defined interestingness function (relevant to her or his workflow and dataset) is a natural way to make best use of limited processing resources in big data workflows, such as HITL systems.




We have shown how a top-$K$ ranked stream-processing workflow model allows proactive tier placement. Approach contrasts over the many of the reactive approaches adopted in the existing literature: our approach does not require monitoring of the application -- and instead allows us to predict and budget for the optimal tier \emph{a priori}.

Considering the document stream as randomly sorted sequence of ranked documents resembles the \emph{Secretary Hiring Problem}, and we can adapt its solutions for hierarchical storage management. The probability of document overwrite/read can be modelled for a class of workflows by introducing an interestingness function. Assuming that documents are ordered randomly (with respect to their interestingness rank) within the stream, our model of document IO holds -- we have demonstrated this with an interestingness trace from a cell-simulation use case. 

We have presented strategies for hierarchical storage management under the top-$K$ workflow. We include case studies of parameter optimization for scenarios with different combinations of cloud storage platforms. The simplicity of the analytic solution makes the approach readily adoptable in production systems, where other domain-specific interestingness functions are defined, work on which we are progressing.

To the authors' knowledge there exists no other work which investigates either top-$K$ workloads, or the secretary hiring problem in the context of cloud resource allocation, or hierarchical (or related) storage systems.

\section{FUTURE WORK}
We are extending the evaluation in Section~\ref{sec:sysbio} to production systems; and the inclusion of a microscopy streaming case study: in this context analysis is severely limited by the cost of storing and analyzing documents. 
Being able to automatically at acquisition time (i.e. online) select which documents to store and analyze further 
could greatly increase the value of an experiment for the same total cost. In many cases, interestingness functions can be defined that are low in processing cost and still are able to select the valuable documents. Interestingness functions can favour images of high quality (in focus and having appropriate signal levels), or images containing the object type of interest (e.g. cells of a certain type, or those with high information content).




\bibliographystyle{apalike}
{
\bibliography{My_Library}}

\end{document}